\documentclass[11pt, a4paper, onecolumn, twoside]{article}
\usepackage{graphicx}
\def \ket #1{\vert #1 \rangle}

\def \bra #1{\langle #1 \vert}
\def \bras  #1{\langle #1 }
\begin{document}
\thispagestyle{empty}
\title{ Proposal of an extended t-J Hamiltonian for high-Tc cuprates
from ab initio calculations on embedded clusters.
}
\author{ Carmen J. Calzado\footnote{ On leave from: Departamento de Qu\'{\i}mica F\'{\i}sica.
 Universidad de Sevilla. E-41012. Sevilla. Spain. } $\:$  and Jean-Paul Malrieu
\\Laboratoire de Physique Quantique. IRSAMC.
\\Universit\'e Paul Sabatier, 31062 Toulouse, France.}
\date{}
\maketitle
\begin{abstract}
A series of accurate ab initio calculations on Cu$_p$O$_q$ finite clusters,
properly embedded on the Madelung potential of the infinite lattice, have
been performed in order to determine the local effective interactions
in the  CuO$_2$ planes of La$_{2-x}$Sr$_x$CuO$_4$ compounds. The values of the
first-neighbor interactions, magnetic coupling ($J_{NN}$=125 meV) and hopping 
integral ($t_{NN}$=-555 meV), have been confirmed.
Important additional effects are evidenced, concerning essentially 
the second-neighbor hopping integral $t_{NNN}$=+110meV, the displacement of a
singlet toward an adjacent colinear hole, $h_{SD}^{abc}$=-80 meV, a non-negligible
hole-hole repulsion $V_{NN}-V_{NNN}$=0.8 eV and a strong
anisotropic effect of the presence of an adjacent hole on the values of the
first-neighbor interactions.
The dependence of $J_{NN}$ and $t_{NN}$ on the position of neighbor hole(s) has
been rationalized from the two-band model and checked from a series of 
additional ab initio calculations. An extended t-J model Hamiltonian has been proposed on
the basis of these results. It is argued that the here-proposed three-body effects
may play a role in the charge/spin separation observed in these compounds, that is,
in the formation and dynamic of stripes.
\end{abstract}
\newpage
\section{Introduction}
The insulating cuprates, such as La$_2$CuO$_4$, which are the undoped parent compounds
of the high-T$_c$ superconducting La$_{2-x}$Sr$_x$CuO$_4$, are known to present 
antiferromagnetic couplings between nearest neighbor (NN) copper centered sites 
in the CuO$_2$ planes. Raman and neutron diffraction experiments evaluate this 
coupling to be around $J_{NN}$=130 meV(128$\pm$6 meV\cite{Sulewski,Singh},134$\pm$5
meV\cite{Aeppli, Endoh, Hayden}, respectively). Nevertheless, the corresponding 
simple Heisenberg Hamiltonian does not reproduce entirely the features of the
Raman spectrum \cite{Parkinson,Canali,Roger:89,Gagliano, Dagotto, Nori:92} and
additional effects such as second-neighbor magnetic coupling $J_{NNN}$ and four-spin
cyclic exchange have been invoked \cite{Eroles:99, Lorenzana:99, Honda:93, Sakai:99}.
While an upper bond for $J_{NNN}$ ($|J_{NNN}|\le$9 meV) has been given from Raman experiments\cite{Hayden}, 
the amplitude of the four-spin operator in this kind of compounds is a matter of discussion.  
Previous works have shown that this cyclic operator
corresponds to a fourth-order term in the Hubbard model, scaling as 
$\lambda t_{NN}^4/U^3$, with $\lambda$=40 \cite{Maynau:82a, Maynau:82b} or  $\lambda$=80 
\cite{MacDonald:88, MacDonald:90}, depending on the formal writing of the Hamiltonian.

Regarding the hole-doped material, where the conduction takes place in the CuO$_2$ planes, 
the holes can be seen as centered on copper atoms with large tails on the four neighboring 
oxygen atoms. They move from on site to an adjacent one through the effect of a hopping operator
 of amplitude $t_{NN}$, for which there is no direct experimental evaluation, but values
 around -0.5 eV are considered as reasonable \cite{Emery:88}. 
One of the most widely employed model Hamiltonians used for the interpretation of the properties
 of these materials, through a hole-pairing mechanism, is the 
 so-called $t-J$ model \cite{Zhang:88, Zhang:90} which combines spin coupling and hole
 hopping:
\begin{equation}
H= \sum_{ ab} J_{NN}\cdot (S_a \cdot S_b -1/4) + t_{NN} \cdot (a_a^{\dag}a_b + a_b^{\dag}a_a + s.c.)\delta(n_a+n_b,1)
\end{equation}
The adequacy of such a simple Hamiltonian to incorporate the physics of the problem is
 questionable. Hopping between second-neighbor sites may be non negligible. When one derives 
the $t-J$ Hamiltonian from the Hubbard Hamiltonian, three-site operators moving a
 singlet-coupled electron pair toward the hole appear at second order of perturbation theory, 
scaling as the $J_{NN}$ operator, i.e, as $t_{NN}^2/U$, where $U$ is the on-site Coulombic repulsion. 
The transferability of $J_{NN}$ from the undoped to the doped material is not guaranteed, the presence
 of a neighboring hole may affect the coupling of two adjacent spins. 
The hole-hole repulsion $V_{ij}$ is likely to play a role, influencing the mean distance 
between the holes. Different extensions of the $t-J$ model have been employed in numerical
 simulations, for instance, $t-t'-J$\cite{White, Martins, Tohyama, Kim} or 
$t-J-V$\cite{Gazza,Riera,Han},but the values given to the parameters are rather arbitrary, 
varying widely from one author to the other, the main objective being to exhibit 
qualitative collective effects. Among them the experimental evidence of the occurrence
 of stripes have focused attention in the recent past\cite{Orenstein}.
 
 The goal of the present paper is to bring useful informations regarding the local effective interactions
 in undoped and hole-doped cuprates. To obtain them, the most accurate tools of ab initio quantum
 chemistry will be used. The method consists in considering few-site clusters, properly embedded
 in the field of the periodic environment, and to calculate the low part of the spectrum using the
 exact Hamiltonian, large basis sets and extensive Configuration Interactions. From this spectrum 
it is possible to fix the amplitudes of the effective interactions. The procedure has been
 successfully used to calculate the NN magnetic coupling in a series of perovskites\cite{perovskitas},
 among them La$_2$CuO$_4$ for which a value of $J_{NN}$=130 meV is obtained. Similar calculations of 
the hopping integral in the hole-doped system ($t$=-0.57 eV) have been also 
reported \cite{Calzado:99a, Calzado:99b}.
These calculations concerned symmetrical
 two-site clusters, for which the determination of $J_{NN}$ and $t_{NN}$ is straightforward 
from the two lowest eigenvalues. This is no longer the case when one considers larger clusters 
to extract additional parameters, concerning next-nearest neighbor (NNN) interactions, 
neighboring-hole dependence of $J_{NN}$ and $t_{NN}$, hole-hole repulsions and four-spin cyclic operators.
Their achievement requires some additional mathematical tools such as the Bloch 
definition\cite{Bloch} of effective Hamiltonians and localization procedures
 (for instance, Boys method \cite{Boys}). The methodology will be explained in Section II.
Section III will present the results concerning the four-site square plaquette and three-site linear cluster
with different number of electrons (holes).
Calculations performed on undoped clusters provide the values of the NN, NNN and next NNN magnetic couplings (Scheme I).
Also the amplitude of the four-spin cyclic exchange for this compound has been established.
Hole-doped clusters give informations about the hopping integrals (NN, NNN and next NNN), the 
singlet-displacement operator and the dependence of the first-neighbor interactions 
(hopping integral and magnetic exchange) 
on the number and relative
position of the adjacent holes.
Section V presents a rationalization of the anisotropy of the effect of hole(s) in the vicinity
on the values of $J_{NN}$ and $t_{NN}$
 and  reports additional
 exploratory calculations to evaluate the dependence of the bicentric parameters
 on the hole positions. Finally, Section VI summarizes the results, proposing a refined 
$t-J$ model, and discusses the possible effect of the additional operators on the charge/spin 
distribution on the lattice, with possible consequences for the stripping phenomena.

SCHEME I\\
\begin{figure}[htb]
\vbox to 1.3cm{
\includegraphics{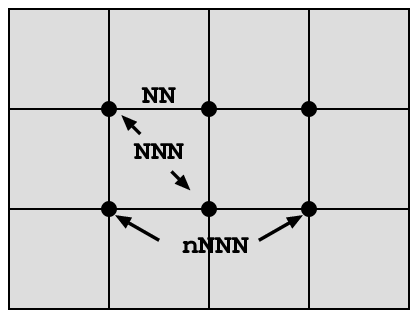}
\vfill}
\end{figure}   
\section{Method}
\subsection{Mapping of a model Hamiltonian on an ab initio effective Hamiltonian}
For such materials the unpaired electrons are essentially located on Cu $dx^2-y^2$ 
in-plane atomic orbitals, with non-negligible delocalization tails on the adjacent oxygen atoms.
 Such Cu-centered orbitals will be labelled $\{a, b, c,..\}$. In the doped material the hole has much
 larger delocalization tails on O $2p$ orbitals, but as shown by Zhang and 
Rice\cite{Zhang:88, Zhang:90} it remains possible to work within a one-band 
model Hamiltonian, the precise nature of its valence orbitals being implicit.
For a finite cluster involving $p$ centers and $n\le p$ unpaired  electrons,
the model Hamiltonian works in a basis of $n$-electron localized determinants
{$\phi_i$}, for which in the Heisenberg and $t-J$ Hamiltonians, the double occupancy
 of the orbitals is prohibited. 
 
 The ab initio calculations handle a large number of atomic
 orbitals, symmetry-adapted molecular orbitals (MOs) and expansions of the wave-functions 
on millions of determinants. Nevertheless, it is possible to construct from these calculations
 ab initio effective Hamiltonians which are in one-to-one correspondence with the model 
Hamiltonians. For an undoped cluster involving $p$ Cu atoms ($p$ sites), it is possible 
first to obtain from variational calculations a set of molecular
 orbitals containing doubly occupied MOs ($core$), unoccupied MOs ($virtual\:\:MOs$) and
 $p$ MOs with essentially single occupation, which define the ab initio 
one-electron valence space, $\{\varphi_i\}$ in Figure 1. A unitary localizing transformation of the $p$ 
symmetry-adapted orbitals will provide equivalent localized orbitals $\{a', b', c'..\}$ which
 can be seen as in strict one-to-one correspondence with the implicit valence orbitals of
 the model Hamiltonian. Fixing a double occupancy of the $core$ orbitals, and putting $n$ electrons
 in these orbitals, avoiding their double occupation,  
localized neutral determinants, $\{\phi_{i,\:loc}'\}$, are obtained which are in correspondence 
with the $n$-electron basis of the model Hamiltonian. These determinants define a model space
(of projector $P_S$, $P_S=\sum \ket{\phi_{i,\:loc}'}\bra{\phi_{i,\:loc}'}$) for the ab initio
calculations. Let be $N_S$ the dimension
of that space. The information obtained by the most refined ab initio calculations will
 be extracted according to the theory of the effective Hamiltonian proposed by Bloch\cite{Bloch}.
 When one knows the $N_S$ eigenstates $\Psi_m$ having the {\it largest
 projections} on the model space (which constitute the target space, stable subspace of $H_{exact}$)
 and their eigenvalues $E_m$,
the effective Hamiltonian is such that its eigenvalues are the
 exact ones, and its eigenvectors are the projections of the exact eigenvectors onto the model
 space:
\begin{equation}
H^{eff}\ket{P_S \Psi_m}=E_m\ket{P_S \Psi_m}, \:\:\: m=1,N_S
\end{equation}
The spectral definition of $H^{eff}$ is:
\begin{equation}
H^{eff}=\sum_m\ket{P_S\Psi_m}E_m\bra{P_S\Psi_m^{\perp}}
\end{equation}
where $\ket{P_S\Psi_m^{\perp}}$ is the biorthogonal transformation of $\ket{P_S\Psi_m}$.
Actually the projections $\ket{P_S\Psi_m}$ of the (orthogonal) states $\Psi_m$ 
have no reason (expect for symmetry reasons) to be orthogonal, they define an overlap matrix $s$:
\begin{equation}
s_{mn}= \bras{P_S\Psi_m}\ket{P_S\Psi_n}
\end{equation}
and the biorthogonal vectors are defined by
\begin{equation}
\ket{P_S\Psi_m^{\perp}} = s^{-1} \ket{P_S\Psi_m}
\end{equation} 
The values of the norms of the projections, i.e. the diagonal elements of $s$ matrix, give
an indication on the quality of the description of these states by the truncated space $S$. 
The model space and the exact eigenstates must be in strong correspondence, i.e. one must
choose both spaces so that the vectors $\Psi_m$ have the largest projections on the model space. 

Then one may express this Hamiltonian in the basis of the localized determinants
 $\{\phi'_{i,loc}\}$ written in terms of the orbitals $\{a', b', c',.\}$ and the matrix elements
 \begin{equation}
\bra{\phi'_i}H^{eff}\ket{\phi'_j}=\sum_m\bras{\phi'_i}\ket{P_S\Psi_m}E_m\bras{P_S\Psi_m^{\perp}}\ket{\phi'_j} 
\end{equation}
can be identified to the matrix elements $\bra{\phi_i}H\ket{\phi_j}$ of the model Hamiltonian. 
In principle the effective Hamiltonians may be non-hermitian but the hermitization
 is straightforward \cite{desCloizeaux}. The comparison between the ab initio effective
 Hamiltonian and the model Hamiltonian fixes the amplitudes of the integrals appearing
 in the latter and allows one to verify whether non-negligible additional interactions are
 not present. Figure 1 summarizes the whole process. Changing the size of the cluster, 
 for instance going from a two-center/one-electron
problem to a three-center/two-electron one, one may check the consistency of the
procedure and the transferability of the effective interactions.

 \subsection{Computational details}
The widely used embedded cluster technique has been employed to model the system.
Finite clusters of the type Cu$_p$O$_q$ (Cu$_3$O$_{10}$ and Cu$_4$O$_{12}$, Figure 2)
have been selected, 
where the $q$ oxygen atoms are the first in-plane neighbors of the $p$ Cu atoms. 
(Previous calculations have shown that
the explicit involvement of the out of plane oxygen atoms does not change the values
of the in-plane interactions \cite{Calzado:99a, Calzado:99b}).
The first-shell of atoms surrounding the cluster have been  
replaced by formal charges with pseudopotentials, in order
to mimic the coulombic and exclusion effects.  
The rest of the lattice
has been modeled by means of point charges, which values have been fixed according 
to Evjen's method\cite{Evjen}, and which correctly represent the Madelung potential of the crystal\cite{Martin}.
This is a simplified approach in comparison with more elaborate methods proposed in the literature\cite{Embedding}.

The ten most internal electrons of Cu atoms have
been represented by effective core pseudopotentials, the valence electrons being
treated explicitly with triple-$zeta$ basis sets. A double-$zeta$ basis set has
been used for oxygen atoms (preliminary calculations on bicentric clusters
have shown that the inclusion of polarization
functions on the bridging oxygen atoms has not an important effect
neither on the magnetic coupling nor on the hopping integral)\cite{Detalles}.

For undoped clusters, 
the restricted-spin open-shell Hartree-Fock calculations
variationaly define the singly occupied magnetic orbitals. These orbitals 
define a minimal valence complete active space (CAS). From this space it is possible
to calculate the spectrum through a difference dedicated Configuration Interaction (DDCI)
procedure \cite{DDCI} which implies all the simple and double excitations on the 
top of this CAS, except the double excitations from the $core$ to the $virtual$ orbitals,
which do not contribute to the energy difference at second-order of perturbation theory \cite{DDCI}.

An alternative solution consists in defining an enlarged CAS including the on-bond $2p$
orbital of the bridging oxygen atoms. These ligand-centered orbitals are the
most participating on the intersite spin-exchange and electron transfer processes \cite{Detalles-OMs}. 
Performing all the single excitations on the top of this extended CAS, 
which corresponds to the two-band Hubbard model, one introduces dynamical polarization effects,
 i.e., screening by the non-active electrons, at lower computational cost than the preceding
 computational scheme.

\section{Ab initio calculations on the plaquette and the linear clusters}

As was mentioned above, two different clusters have been used to extract the effective
interactions.
A four-site square cluster (plaquette) of formula Cu$_4$O$_{12}$ has been employed 
in order to determine the first- (NN) and second-neighbor (NNN) interactions, and also the four-spin
cyclic exchange. Third-neighbor interactions (nNNN) has been estimated by means of the
calculations carried out in a linear three-site cluster (Cu$_3$O$_{10}$). Comparing with
the results obtained from the plaquette and previously studied binuclear clusters, it is  
possible to check the dependence of the NN interactions on the
size of the fragment involved in the ab initio calculations.\\

Three fillings of the {\it valence shell } have been considered in order to
evaluate the dependence of these interactions on the hole concentration. Undoped (4centre/4electron 
and 3centre/3electron problems), one hole-doped (4c/3e and 3c/2e) and two hole-doped (4c/2e and
3c/1e) situations have been analyzed.
From the systems with two holes in the valence shells, it
is possible to extract the amplitude of the hole-hole repulsions, an important magnitude for the study of 
the hole pairing mechanism. 
It is worth to notice that the here-referred hole-dopings are not in correspondence with the
total doping of the lattice, induced by the replacement of La$^{+3}$ by Sr$^{+2}$. 
A change in the occupation of the valence shell of these small clusters
just induces a $local$ hole doping, which
provides informations about the $local$ modifications of the effective parameters. 

\subsection{The localization process}
In the plaquette, the four symmetry-adapted valence
orbitals belong to the $A_{1g}$, $E_u$ and $B_{1g}$ representations in the $D_{4h}$ group. 
The localizing unitary transformation is straightforward since:
\begin{eqnarray}
a_{1g}&=&(a+b+c+d)/2 \\
e_{u(1)}&=&(a+b-c-d)/2 \\
e_{u(2)}&=&(a-b-c+d)/2 \\
b_{1g}&=&(a-b+c-d)/2 
\end{eqnarray}
Figure 3 pictures one of these four localized valence orbitals for undoped and for the
doped plaquettes, showing the strong localization of the magnetic orbitals and the $d-p$
 hybridization occurring in the hole-doped systems\cite{Calzado:99a, Calzado:99b,Uchida}.

In the linear cluster the localizing transformation of the three magnetic orbitals
is no longer imposed by the symmetry. Two of these orbitals ($\varphi_g, \varphi_g'$)
belong to the $A_g$ irreducible representation and the other one ($\varphi_u$) to the 
$B_{1u}$ symmetry. The rotation $U$, which transforms the $\{ \varphi_g, \varphi_g', \varphi_u \}$ 
into the localized set $\{ a',b',c'\}$ has been performed according to the Boys
criterion \cite{Boys}, which maximizes the distance between the centroids of the orbitals.
An alternative localization criterion, the minimization of the direct exchange
integral $K_{ac}$, leads to the same rotation.

\subsection{The magnetic interactions}

Table 1 reports the results obtained from the plaquette and
the linear clusters for the magnetic coupling,
involving different fillings of the valence shell. 
For undoped cluster, the first-neighbor magnetic coupling $J_{NN}$ is $\sim$ 125 meV. The calculated value
is independent on the size of the considered cluster and it is in agreement
with both the previously determined
$J_{NN}$ in binuclear cluster and the estimations from Raman  and neutron diffraction
experiments (128$\pm$6 meV\cite{Sulewski,Singh}, 134$\pm$5
meV\cite{Aeppli, Endoh, Hayden}, respectively). 
The coupling between second-neighbors $J_{NNN}$ is also antiferromagnetic, with a value of 
$J_{NNN}$=6.5 meV in accordance with the experimental upper limit $|J_{NNN}|\le$9 meV \cite{Hayden}. 
A negligible antiferromagnetic coupling has been found between Cu atoms placed at a 2R distance
(third-neighbors, next NNN): $J_{nNNN}$=1 meV.

Finally, the calculations on the plaquette provide a non-negligible value for the four-spin
cyclic exchange of $K$= 14 meV. The four-body operator can be written as:
\begin{equation}
H_K= K \sum_{<ijkl>} [  (S_i\cdot S_j)(S_k\cdot S_l)+(S_i\cdot S_l)(S_j\cdot S_k)-(S_i\cdot S_k)(S_j\cdot S_l)] 
\end{equation}
This operator produces the cyclic permutation of the four spins on the plaquette (Scheme II) plus ordinary 
two-spin exchanges of all the pairs of spins of the plaquette including those of the diagonals.
(A detailed discussion has been reported elsewhere \cite{4cuerpos}.)
Its value is
somewhat smaller than some estimations used in numerical simulations
\cite{Honda:93,Lorenzana:99,Schmidt}, but 
larger than the critical value, $(K/J_{NN})_c$=0.05 $\pm$0.04, estimated by
Sakai and Hasegawa \cite{Sakai:99} for the appearance of a magnetization plateau at half the 
saturation value in the $S=\frac{1}{2}$ antiferromagnetic spin ladders.

SCHEME II\\
\begin{figure}[htb]
\vbox to 0.5cm{
\includegraphics{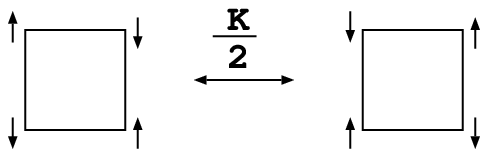}
\vfill}
\end{figure}   

When a hole is introduced in the cluster, the NN magnetic coupling is influenced by its 
presence, but in different directions depending on the relative position of this hole. Thus
a hole in a colinear position to the two NN spins {\it increases} the coupling
between these two spins (Scheme IIIa).
However the NN magnetic coupling $diminishes$ if the hole is placed in a position perpendicular
to the bond (Scheme IIIb).
The same trend is observed in the plaquette when a second-hole is introduced, $J_{NN}$ being 94 meV,
to be compared with $J_{NN}$= 104 meV in presence of 1 hole and $J_{NN}$=125 meV for undoped systems.

SCHEME III\\
\begin{figure}[htb]
\vbox to 1.4cm{
\includegraphics{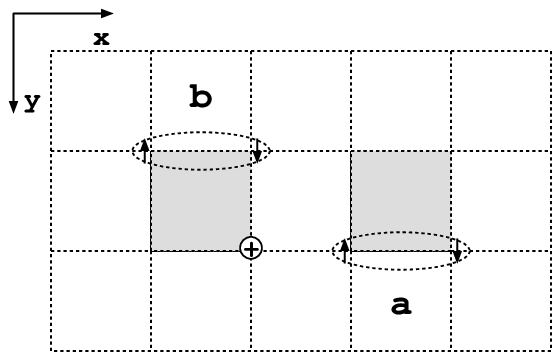}
\vfill}
\end{figure}   

\subsection{Hopping integrals}
As in the case of magnetic interactions, the first-neighbor hopping integral ($t_{NN}$) is also
independent on the size of the cluster (Table 2), a value of -558 meV has been found in 1hole-doped linear
cluster, -552 meV in 1hole-doped plaquette and -555 meV in previously reported binuclear
cluster\cite{Calzado:99a,Calzado:99b}. This value is in accordance with
a generally accepted value of -500 meV for these compounds\cite{Emery:88}.

An evaluation of second- and third-neighbor hopping integrals has also been possible. The NNN hopping
integral $t_{NNN}$=+112 meV is unexpectedly large. The sign is in agreement with the negative overlap
of the active orbitals placed at a $\sqrt 2 R$ distance, but its magnitude is large due to 
through bond processes, which involve the oxygen atoms.
On the basis of the perturbation theory, there are two contributions to the $t_{NNN}$ hopping
integral.  
One corresponds to a  
a third-order contribution, 
scaling as $\frac{-t_{pp} (t_{pd})^2}{\Delta E_{CT}^2}$, 
where $t_{pd}$ is the  hopping integral between the O $2p$ 
and the Cu $3d$ orbitals, $t_{pp}$ is the hopping integral between the O $2p$ orbitals and 
$\Delta E_{CT}$ is the O $2p\rightarrow$ Cu $3d$ charge-transfer
 excitation energy (Scheme IV).
Since $t_{pp}$, $t_{pd}$ and $\Delta E_{CT}$ are negative quantities, the third-order contribution
results in a positive magnitude.
There exists an alternative pathway corresponding to a fourth-order contribution, 
scaling as $\frac{t_{pd}^4}{\Delta E_{CT}^3}$, with 
opposite sign.
Since the third-order contribution is expected to be larger than the fourth-order one, the
resulting sign of $t_{NNN}$ is positive.

A moderate amplitude of the hopping integral between third-neighbors (distance 2R) has been extracted
from the calculations on the linear cluster. The extension of this coupling is controlled by through-bond
interactions, but in contrast with the plaquette, only fourth-order process can be written here,
smaller in magnitude and with an opposite sign with respect to $t_{NNN}$.

SCHEME IV\\
\begin{figure}[htb]
\vbox to 5.0cm{
\includegraphics{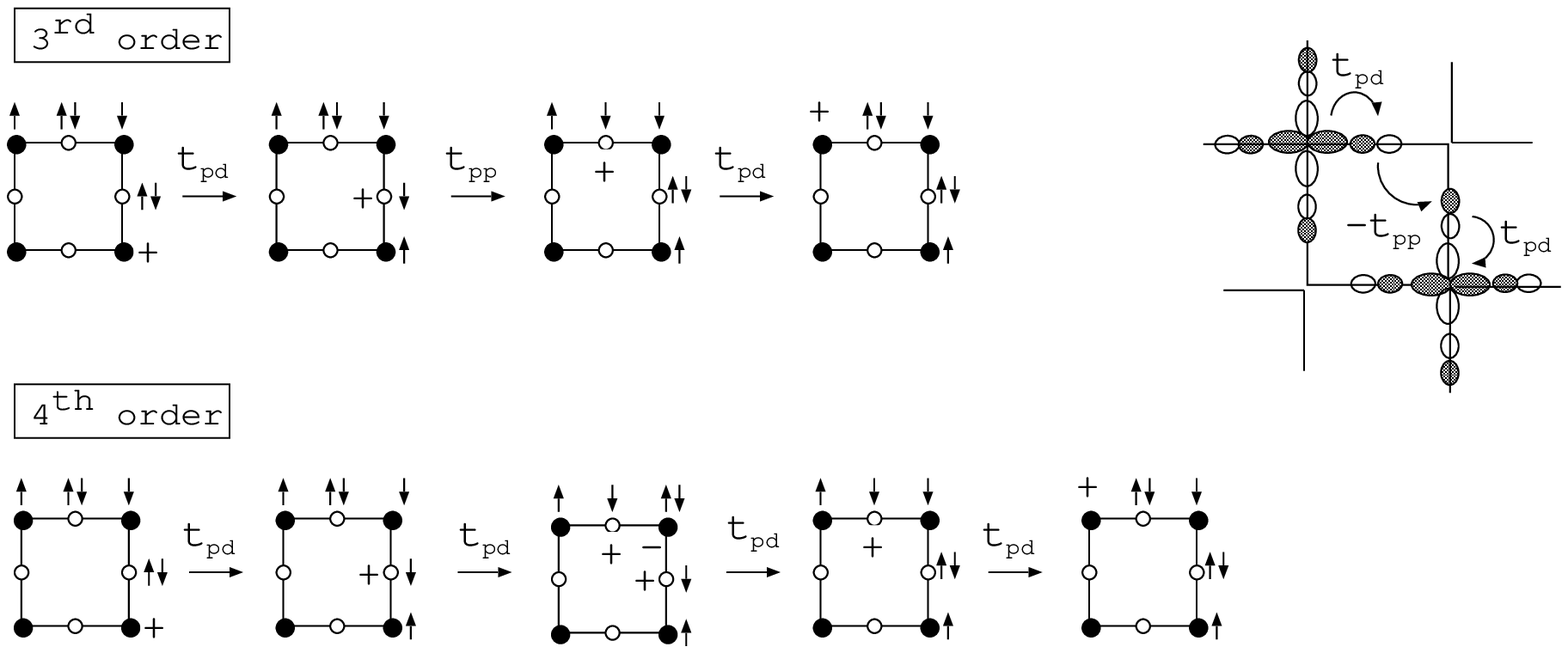}
\vfill}
\end{figure}   

When an additional hole is introduced in the system, the NN hopping integral is unchanged
when the hole is placed in a perpendicular position to the bond ($t_{NN}$=-558 meV 
in the 2hole-doped plaquette, Scheme Va), but its absolute value is augmented
when the hole is placed in a colinear position to
the bond ($t_{NN}$=-600 meV in the 2hole-doped linear cluster, Scheme Vb).  

SCHEME V\\
\begin{figure}[htb]
\vbox to 1.8cm{
\includegraphics{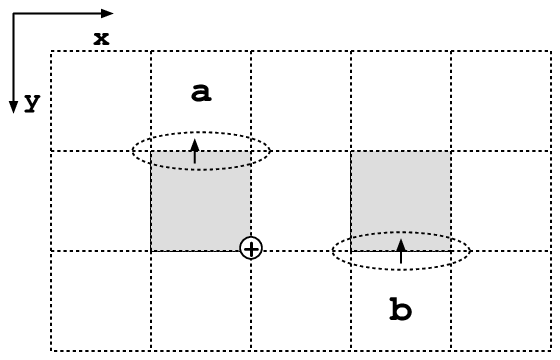}
\vfill}
\end{figure}   

The presence of the extra hole does not influence significantly neither 
the second-neighbor ($t_{NNN}$= +130 meV to
be compared with +112 meV in absence of this additional hole) nor
the third-neighbor ($t_{nNNN}$=-36 meV versus -47 meV) hopping integrals. 

\subsection{Singlet displacement operator}
An additional information coming from these calculations concerns the singlet-displacement operator(Table 3).
It is a three-site/two-electron operator, which moves the pair of electron, coupled in a singlet, 
toward a hole placed in a neighbor position. Thus, a singlet on sites $a$ 
and $b$, $c$ containing a hole,  is displaced to the positions $b$ and $c$,
 the hole being in $a$ (Scheme VI):\\
\\
SCHEME VI\\
\begin{figure}[htb]
\vbox to 1.2cm{
\includegraphics{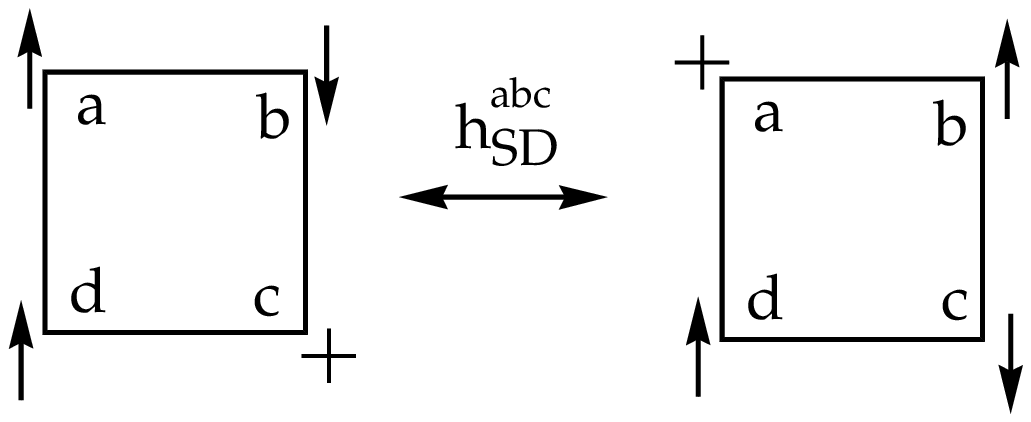}
\vfill}
\end{figure}   

It can be written as:
\begin{equation}
h_{SD}^{abc}\left[ \ket{a\bar{b}-\bar{a}b}\bra{b\bar{c}-\bar{b}c} + 
\ket{b\bar{c}-\bar{b}c}\bra{a\bar{b}-\bar{a}b}\right]\delta(n_a+n_b+n_c,2)
\end{equation}
where $\delta(n_a+n_b+n_c,2)$ controls the fact that the three centers bear only 
two electrons, i.e. that the singlet can only move toward a hole. 
As in the precedent parameters, the amplitude $h_{SD}^{abc}$ depends on the relative
position of the hole. If the singlet moves to a neighbor bond in the plaquette (i.e., the
hole is placed in a perpendicular position to the singlet bond), the value is 
$h_{SD}^{abc}$= -41 meV while it goes to 
 $h_{SD}^{abc}$=-80 meV if the hole is on the same axis than the singlet. 

The presence of a second hole in the plaquette does not affect the amplitude of
the singlet-displacement ($h_{SD}^{abc}$= -37 meV).

There exists also a small similar operator involving the diagonal of the square
 ($h_{SD}^{cbd}$= 9 meV):
\begin{equation}
h_{SD}^{cbd}\left[ \ket{b\bar{c}-\bar{b}c}\bra{b\bar{d}-\bar{b}d} +
\ket{b\bar{d}-\bar{b}d} \bra{b\bar{c}-\bar{b}c}\right]\delta(n_b+n_c+n_d,2)
\end{equation}
moving the electrons as shown in Scheme VII.
The value of this operator in presence of a second hole is very close to the preceding value:
 $h_{SD}^{cbd}$= 14 meV.

SCHEME VII\\
\begin{figure}[htb]
\vbox to 1.5cm{
\includegraphics{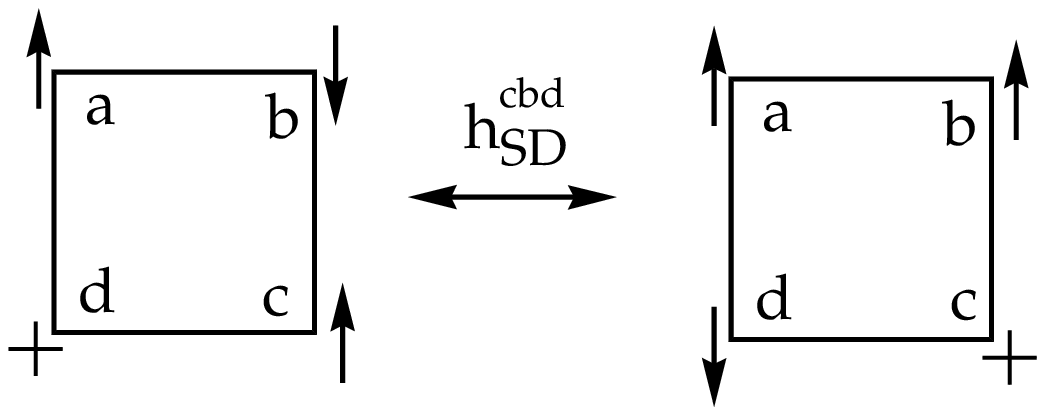}
\vfill}
\end{figure}   

\subsection{Hole-hole repulsions}
The absolute value of the hole-hole repulsion is not accessible, but
 the effective Hamiltonian gives the difference between two situations (Table 4).
 From the plaquette, we can extract the relative stability of two holes placed in NNN positions
with respect to two holes in NN:
\begin{equation}
V_{NN}-V_{NNN}= V_{ab}-V_{ac}= 0.98 eV
\end{equation}

This is significantly larger than the values usually accepted for simulations \cite{Gazza,Riera,Han}. One 
should stress on the fact that this value takes into account the dynamical repolarization
effects of all the atoms explicitly treated in the calculation, i.e., the screening by
the twelve in-plane oxygen atoms linked to the four Cu atoms of the plaquette. It misses
the polarization of the rest of the environment. Taking into account the polarization
of a large surrounding shell, with a value of the polarizability of the O$^=$ ion,
$\alpha{_O}=1.30$ \AA${^3}$ (obtained from a series of finite-field ab initio  calculations on
an embedded $CuO_4$ cluster), in agreement with experimental estimates \cite{polarizabilidad},
one diminishes the hole-hole repulsion difference to
 $V_{NN}-V_{NNN}$=0.80 eV which remains a rather large value.
From the linear cluster, it is possible to estimate the 
energy gain obtained when placing two holes in nNNN positions (distance 2R) instead of adjacent 
positions (distance R):
$V_{NN}-V_{nNNN}= V_{ab}-V_{ac}$= 1.77 eV, with a final value of  $V_{NN}-V_{nNNN}$=1.47 eV  
once the environment polarization effects
have been taken into account. 

These values of {\it differences} between hole-hole repulsions may seem very large and most
of the calculations introducing this repulsion in a t-J-V model use to take smaller values
($J < V < 4J$)\cite{Gazza,Riera,Han} when trying to exhibit hole-pairing phenomenon. The above calculated
values are smaller than the corresponding electrostatic quantities calculated in a point charge
approximation: $V_{NN}-V_{NNN}$=1.12 eV and  $V_{NN}-V_{nNNN}$=1.90 eV. The delocalization of the holes
on the oxygen atoms should result in larger repulsions, especially for $V_{NN}$ since the two holes
share an oxygen atom, and our values, which exhibit a significant screening, do not seem unrealistic.

\section{Interpretation of the hole dependence of the one bond $J_{NN}$ and $t_{NN}$ parameters}
The two preceding sections indicate that the presence of holes in the immediate vicinity 
of a bond may affect the values of the spin coupling and of the hopping integral on this bond,
 and that this effect is anisotropic, i.e., does not only depend on the minimal distance of 
the hole to the atoms of the bond. Hence it seems important to rationalize these effects 
if possible.

 \subsection{Rationalization from the two-band model}
{\it The hopping integral.}\\
The rationalization of this anisotropy is possible in terms of the two-band model. 
The $t$ hopping integral results from a second order effect (Scheme VIII, where the open circle represents
a bridge oxygen atom and closed circles correspond to copper atoms):

SCHEME VIII\\
\begin{figure}[htb]
\vbox to 1.3cm{
\includegraphics{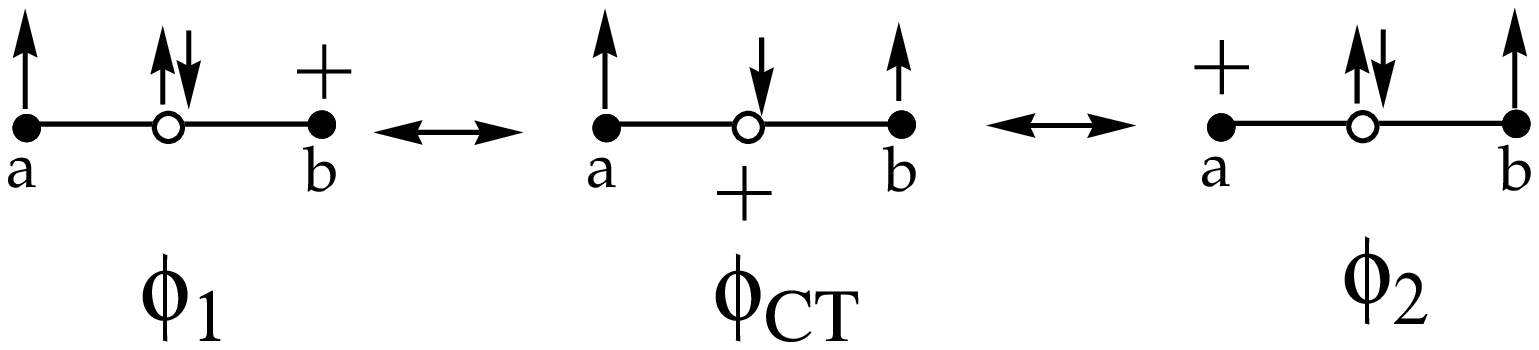}
\vfill}
\end{figure}  
\\
$t$ scaling as 
\begin{equation}
t \sim \frac{t_{pd}^2}{\Delta E_{CT}} 
\end{equation}
where $\Delta E_{CT}$ is the $2p$ O to $3d$ Cu charge transfer excitation energy.
 In presence of an additional hole, the energies of the model space determinants, 
$\phi_1$ and $\phi_2$, and of the intermediate charge transfer state, $\phi_{CT}$, will
 be modified. The model space determinant energies are no longer degenerate and one
 shall take their mean energy as the zero-order energy. Let consider now the effect on
 the bond directed along $x$ of an adjacent hole placed either on the $y$ direction, as 
occurs in the plaquette, or on the $x$ axis, as in the linear cluster (schemes Va and Vb).
In situation (Va) the zero-order mean electrostatic energy with the hole is 
$\delta E_0=\frac{\sqrt{2}+1}{2\sqrt{2}R}=\frac{0.85}{R}$.
 In the corresponding charge transfer state the mean electrostatic energy with 
the hole is $\delta E_{CT}=\frac{2}{\sqrt{5}R}=\frac{0.89}{R}$, i.e., the excitation 
energy is increased by a small quantity:
\begin{equation}
\Delta E_{CT}'=\Delta E_{CT}+\delta E_{CT}-\delta E_0=\Delta E_{CT}+\frac{0.04}{R}
\end{equation}
which should diminish slightly the absolute value of the hopping integral. 
In contrast, for situation (Vb), $\delta E_0=\frac{3}{4R}$ and $\delta E_{CT}=\frac{2}{3R}$
hence $\Delta E_{CT}'=\Delta E_{CT}-\frac{0.08}{R}$ which should increase the absolute 
value of $t$, in agreement with our calculation. 

This analysis is quite rudimentary, 
it neglects the possible effects of the polarization of the orbitals on the $t_{pd}$
 integrals and rests on a crude evaluation of the electrostatic effects on the denominators 
 (holes considered as centered on Cu sites, neglecting their oxygen character),
 but it seems to agree with the computed trends.
 \\
 \\
{\it The magnetic coupling.}\\
The same kind of analysis may be tempted for the $J_{NN}$ coupling constant, which is now a fourth-order
quantity (Scheme IX):
\begin{equation}
J_{NN} \sim\frac {-2t_{pd}^4} {U (\Delta E_{CT})^2}
\end{equation}
which involves intermediate situations of opposite polarities, but with the same
energy: $\Delta E_{CT1}=\Delta E_{CT2}= \Delta E_{CT}$. In the undoped system, $U_1$=$U_2$=$U$,
which corresponds to the Coulombic repulsion of two electrons placed in the same $3d$ orbital.

SCHEME IX\\
\begin{figure}[htb]
\vbox to 5.0cm{
\includegraphics{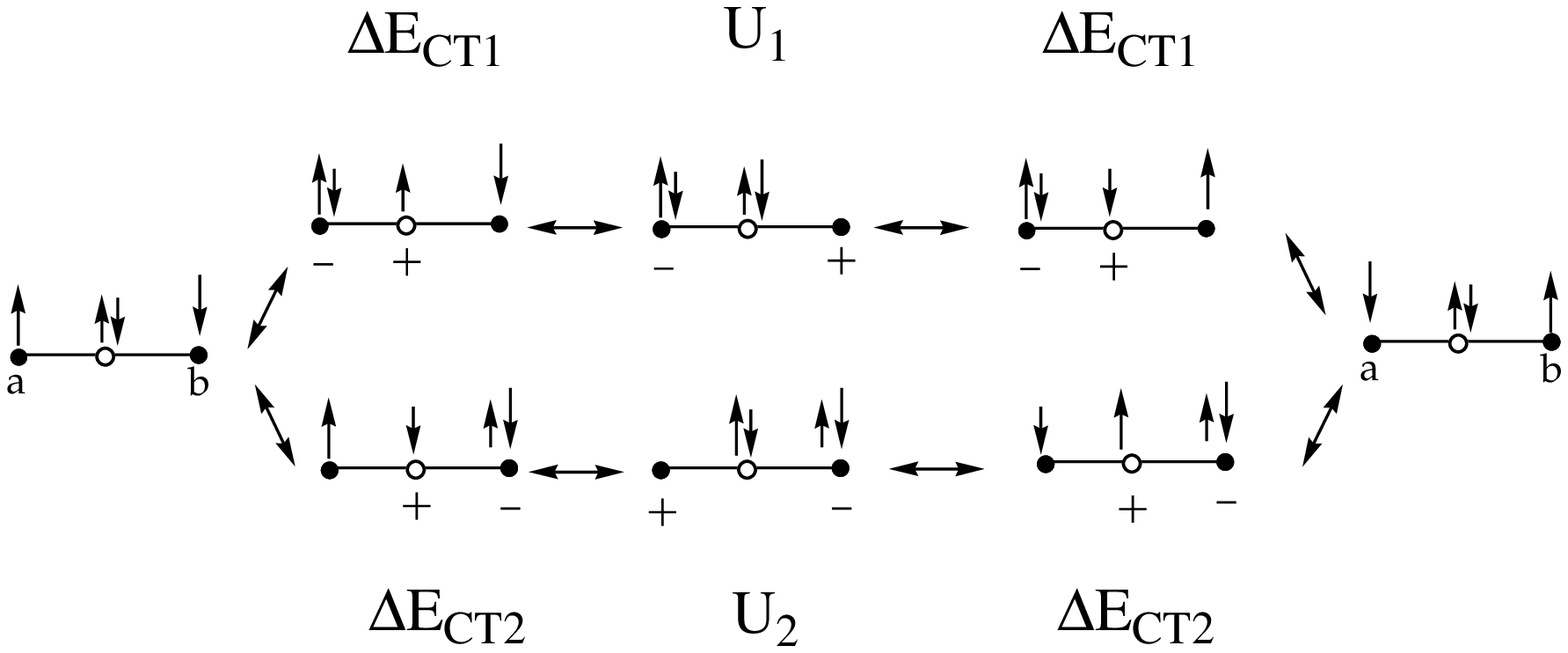}
\vfill}
\end{figure}  
\\
Adding an external hole will not modify the electrostatic zero-order energy, but will affect the 
energies of all intermediate states, so that the coupling becomes:\\
$\frac{-t_{pd}^4}{\Delta E_{CT1}^2}\cdot\frac{1}{U_1}$ (for the top pathway) $+$
$\frac{-t_{pd}^4}{\Delta E_{CT2}^2}\cdot\frac{1}{U_2}$ (for the bottom one).\\
Notice that while $\Delta E_{CT1}=\Delta E_{CT}+\delta E_{CT1}$
 and $\Delta E_{CT2}=\Delta E_{CT}+\delta E_{CT2}$ have no reason to be related, 
$U_1=U+\delta U$ and $U_2=U-\delta U$  whatever the outer charge distribution. 
The effective coupling is therefore:
\begin{equation}
-t_{pd}^4\left(\frac{1}{(\Delta E_{CT}+\delta E_{CT1})^2}\cdot\frac{1}{U+\delta U} + \frac{1}{(\Delta E_{CT}+\delta E_{CT2})^2}\cdot\frac{1}{U-\delta U}\right)
\end{equation}

To the first order in $\frac{\delta U}{U}$ and $\frac{\delta E_{CT}}{\Delta E_{CT}}$
 developments, one gets:
\begin{equation}
J_{NN}= \frac{-2t_{pd}^4}{U(\Delta E_{CT})^2} \left(1-\frac{\delta E_{CT1}}{\Delta E_{CT}}-\frac{\delta E_{CT2}}{\Delta E_{CT}}-\frac{\delta U}{U}+\frac{\delta U}{U}\right)+{\cal O}(2)
\end{equation}

The effect of the additional hole(s) will go through their electrostatic
 interaction in the charge transfer states. For the perpendicular situation (Scheme X) 
$\delta E_{CT1}=\frac{0.187}{R}$ and $\delta E_{CT2}=\frac{-0.105}{R}$. Hence, 
$\delta E_{CT1}+\delta E_{CT2}=\frac{0.082}{R}$ and $J_{NN}$ is diminished by the presence of the adjacent
hole in the $y$ direction, as observed in the corresponding ab initio calculations
 ($\Delta J_{NN}/J_{NN}$=-17$\%$).\\ 
\\
SCHEME X\\
\begin{figure}[htb]
\vbox to 8.0cm{
\includegraphics{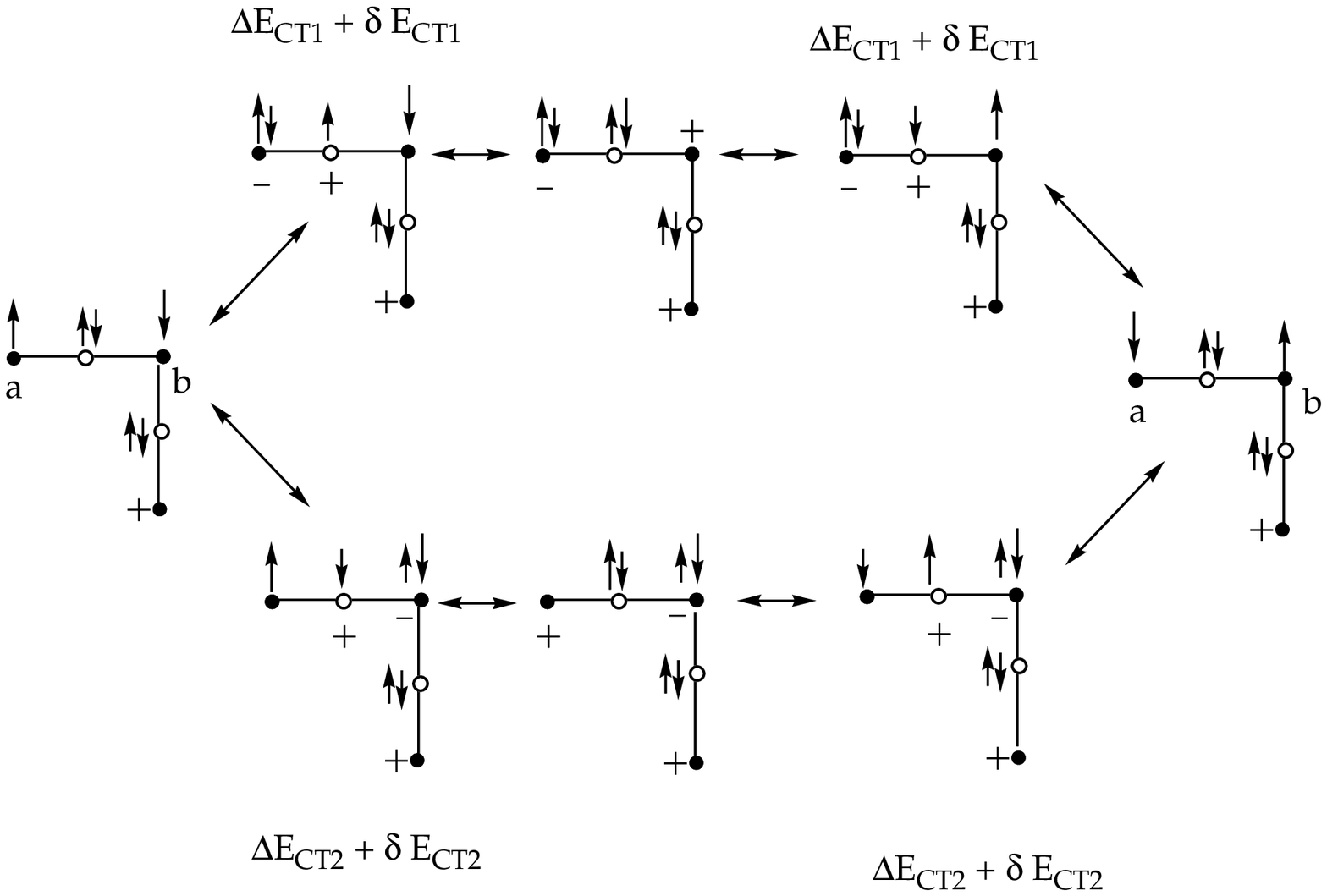}
\vfill}
\end{figure}  
 
 Oppositely, in the linear situation (Scheme XI) the quantities
 are: $\delta E_{CT1}=\frac{1}{6R}$ and $\delta E_{CT2}=\frac{-1}{3R}$, so
 $\delta E_{CT1}+\delta E_{CT2}=\frac{-1}{6R}$
 is a negative quantity, hence the coupling constant should be increased. This is actually 
observed since $\Delta J_{NN}/J_{NN}$=+25$\%$. The variation of $J_{NN}$ in the model is of the right sign
 and its amplitude is larger than for the perpendicular orientation.  \\
 \\
SCHEME XI\\
\begin{figure}[htb]
\vbox to 6.0cm{
\includegraphics{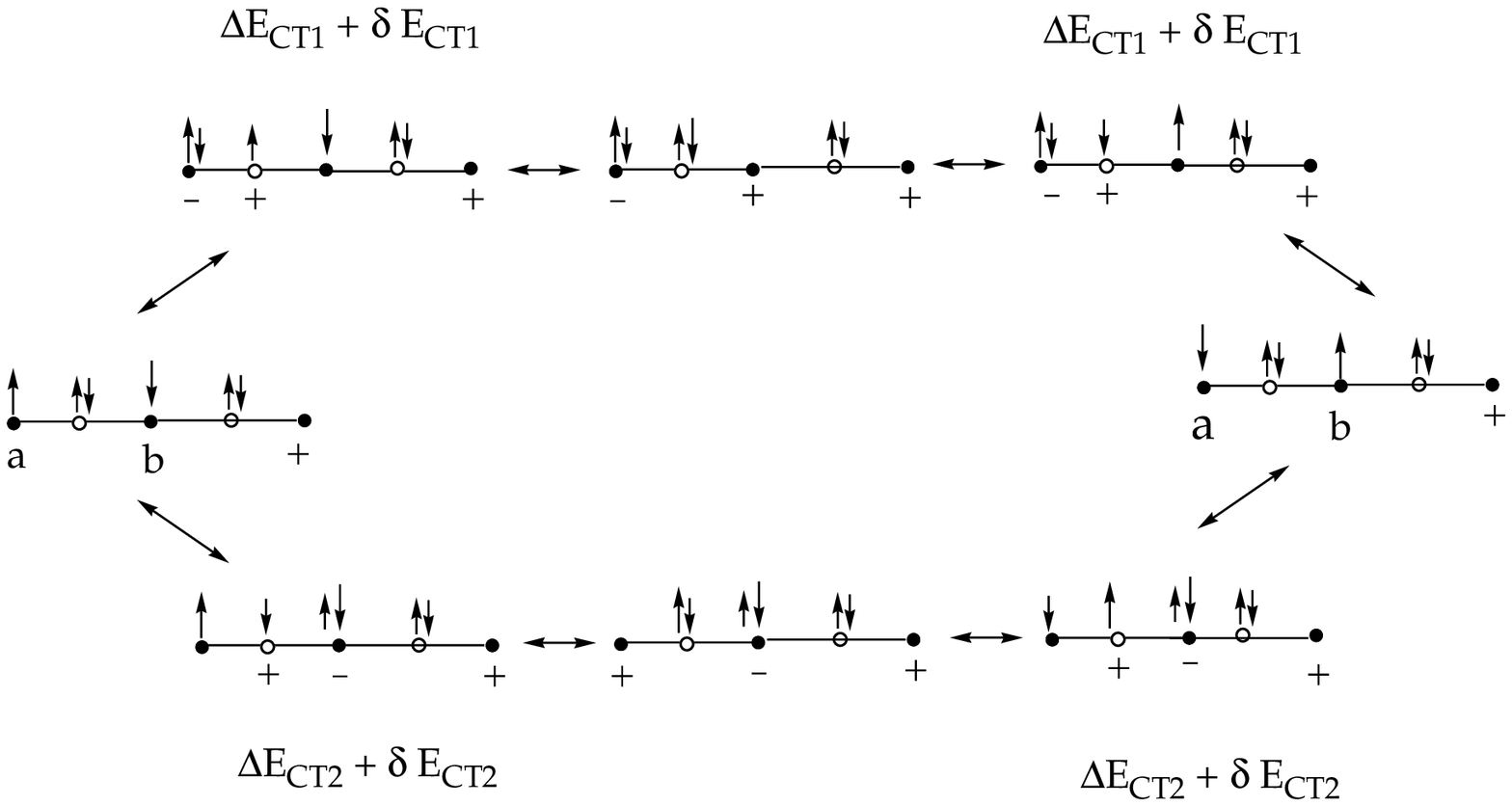}
\vfill}
\end{figure}  
\\
\\
{\it The singlet-displacement operator.}\\
The same kind of rationalization applies to the two-electron/three-centre operator $h_{SD}^{abc}$.
In the one-band model this effect scales as $t^2/U'$ (Scheme XII):\\
\\
SCHEME XII \\
\begin{figure}[htb]
\vbox to 1.0cm{
\includegraphics{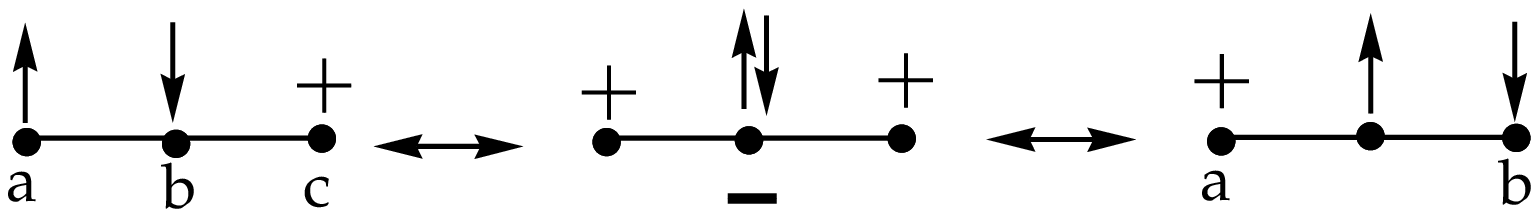}
\vfill}
\end{figure}  
\\
where $U'=U-2V_{ab}+V_{ac}$, $V_{ij}$ being the electrostatic interaction between the charges
(with respect to the basic undoped distribution). The denominator will be smaller for the
 linear configuration than for the perpendicular one since $V_{ac}$ is larger 
($1/\sqrt{2}R$ instead of $1/2R$). This may explain the difference between the 
amplitudes of $h_{SD}^{abc}$ found in the plaquette and in the linear cluster (Table III).
The more realistic two-band model introduces an
 alternative mechanism, since there exist two paths to go from $a\bar{b}$ to $b\bar{c}$, 
cf Scheme XIII:\\
SCHEME XIII \\
\begin{figure}[htb]
\vbox to 5.0cm{
\includegraphics{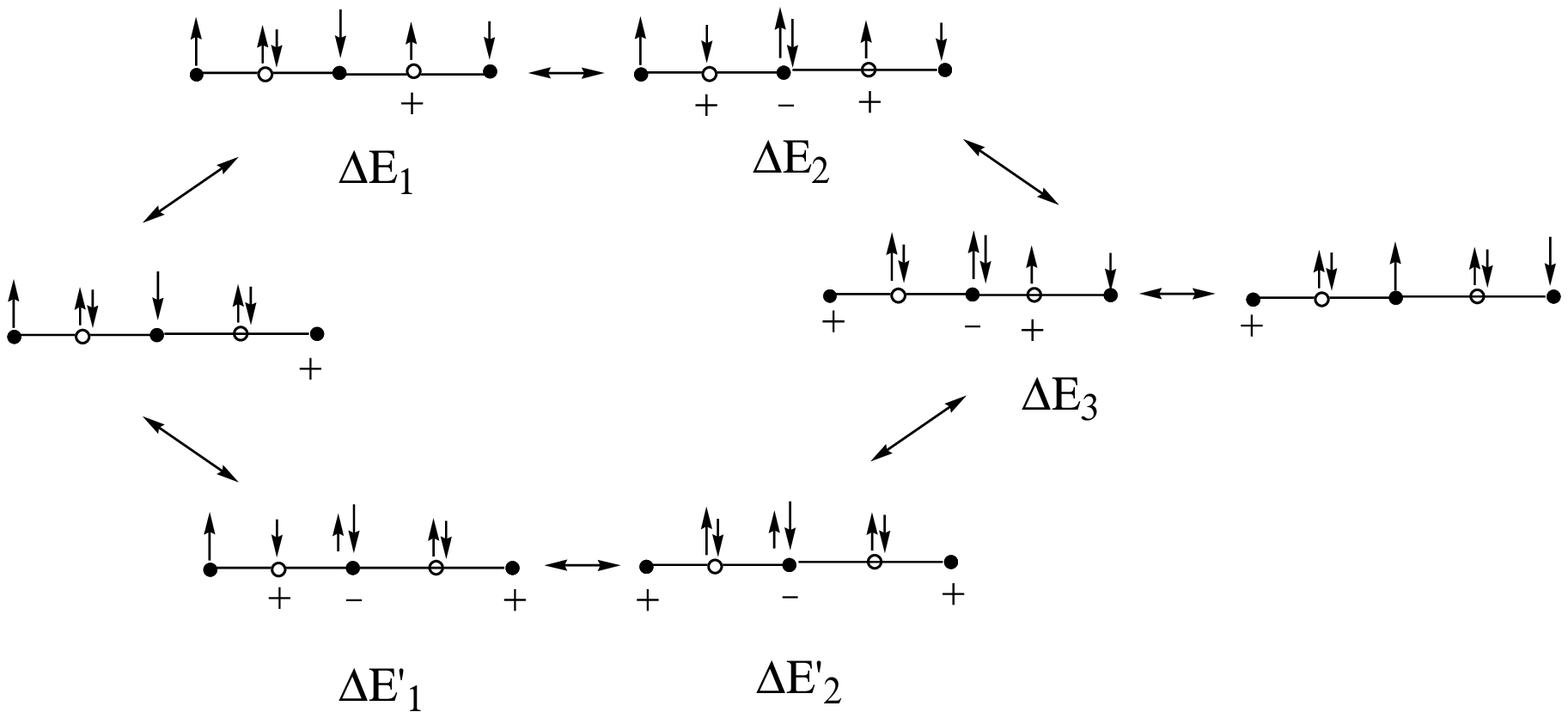}
\vfill}
\end{figure}  
\\
The second path corresponds to the process appearing in the one-band model, which is
 not the case for the first one. Explicitly, the excitation energies are:
\begin{eqnarray}
\Delta E_1&=& E_d - E_p-U_p= \Delta E_{CT}- U_d + \frac{2}{R} \\
\Delta E_2&=& 2E_d-2E_p+U_p-2U_p-\frac{4}{R}+V_{I,II}=2\Delta E_{CT} - U_d + V_{I,II}\\
\Delta E_3&=& E_d - E_p-U_p+U_d-\frac{3}{R}+V_{a,II}= \Delta E_{CT}- \frac{1}{R} + V_{a,II} \\
\Delta E_1'&=&\Delta E_3 \\
\Delta E_2'&=&U_d-\frac{2}{R}+V_{ac} 
\end{eqnarray}
It is likely that $\Delta E_2'> \Delta E_2$, i.e. that the additional path has a larger 
contribution. The overall effect is:
\begin{equation}
 \frac{t_{pd}^4}{\Delta E_1'}\left(\frac{1}{\Delta E_1}\cdot\frac{1}{\Delta E_2}+
\frac{1}{\Delta E_1'}\cdot\frac{1}{\Delta E_2'}\right)
\end{equation}
ie, of the same order of magnitude as $J_{NN}$. Regarding the orientation effect it is clear 
that the denominators are larger for the perpendicular orientation than for the linear one, 
due to the hole-hole repulsions $V_{I,II}$,$V_{1,II}$ and $V_{13}$, and it is expected that
  $h_{SD}^{abc}$ will be larger for the linear orientation, as found in the numerical ab initio 
calculations (-80 meV and -40 meV, respectively).

\subsection {Further exploration of the influence of the hole on $J_{NN}$ and $t_{NN}$}
In order to evaluate this influence, larger clusters would have to be considered,
 which is beyond our computational posibilities (the preceding CI expansions frequently 
reach $10^7$ determinants), and a simpler procedure has to be used. In the present set of
 calculations, a dinuclear $Cu_2O_7$ clusters have been chosen as before, adding one or two
 additional point charges on Cu sites of the environment, which become Cu$^{+3}$ centers, 
in order to grossly mimic the effect of the holes. For first-neighbor holes, this is a rather
 crude approximation, and its relevance has to be tested by comparison with the previously 
reported four- and three-center calculations. Actually, for a perpendicular position of the 
hole $J_{NN}$=115 meV (compared to 104 meV in the plaquette). For a colinear hole 
$J_{NN}$=175 meV, compared to 156 meV in the linear cluster. Hence the anisotropy
 of the first-neighbor hole is qualitatively reproduced. As an additional confirmation 
we have calculated the effect of two holes in the plaquette and we get the same value
 $J_{NN}$=93 meV as for the 2hole-containing plaquette. Table 5 gives the results 
for $J_{NN}$ concerning a series of possibilities with one remote hole, two holes
in different positions, together with the variation of the electrostatic energy in 
the intermediate charge transfer states with respect to the non-doped case,
 expressed in eV. 
Figure 4a shows the correlation between the calculated $J_{NN}$ value with the variation of the
electrostatic energy $\delta E_{CT1}+ \delta E_{CT2}$ in the intermediate charge transfer states.
The correlation is quite satisfactory and suggests the following law in eV:
\begin{equation}
J_{NN}=0.131 -0.053 (\delta E_{CT1}+\delta E_{CT2})
\end{equation}
An analogous fit of the hopping integral dependence on the variation of the electrostatic
 energy in the intermediate charge transfer state $\delta E_{CT}-\delta E_0$ has
 been tempted from the values of $t_{NN}$ (Table 5) calculated on binuclear
 complexes in presence of external hole(s). The correlation is less 
satisfactory (Figure 4b). The distortions of the active orbitals in presence of these 
close holes should be non-negligible and should affect the amplitudes of the hopping 
integral in a rather complex manner. However, it seems simpler to research a linear law 
rather than to produce and handle a dictionary of operator amplitudes considering exhaustively all
possible occurences. We therefore propose the following fit in eV for the first-neighbor hopping
integral:
\begin{equation}
t_{NN}=-0.521+0.187 (\delta E_{CT}-\delta E_0)
\end{equation}

\section{Discussion and Conclusions}
The numerical results obtained here above suggest that the usual t-J or t-J-V model 
Hamiltonins neglect important physical effects, which should be incorporated into an
extended model Hamiltonian. The main deviations from the t-J Hamiltonian are the
following:\\
(i) the inclusion  of the second-neighbor hopping, $t_{NNN}\simeq$+110 meV. The third-neighbor 
interaction $t_{nNNN}\simeq$ -40 meV could tentatively be omitted. 
Recent analysis of angle-resolved photoemission spectroscopy data has shown that both $t_{NNN}$
and $t_{nNNN}$ are necessary for understanding the dispersion and line shape of the spectral function
in the t-J model \cite{Kim}. Tohyama {\it et al.} \cite{Tohyama} have estimated the ratio $t_{NNN}/t_{NN}$
and $t_{nNNN}/t_{NN}$ to be -0.12 and 0.08, respectively, by fitting the tight-binding Fermi surface
to the experimental one in the overdoped sample \cite{Ino} on the assumption that in the overdoped region
the Fermi surface of the tight-binding band is the same as that of the t-t'-t"-J model. 
The values here-proposed are in good agreement with these ratio:   
$t_{NNN}/t_{NN}$ =-0.19 and $t_{nNNN}/t_{NN}$=0.08
but not with the ratio $J/t_{NN}$ proposed by these authors (${J/t_{NN}}_{Tohyama}$=0.4 vs $J/t_{NN}$=0.22).
\\
(ii) the singlet displacement operator $h_{SD}$ has to be taken into account, at least for
the colinear displacement, since $h_{SD}\simeq$ -80 meV.\\
(iii) the hole-hole repulsion appears to be far from negligible. Due to possible screening
effects and electrostatic mean cancellations by the Sr ions, it is certainly reasonable to neglect
the hole-hole repulsions beyond the third-neighbors. Even if the calculated values $V_{NN}-V_{NNN}$=0.8 eV
and $V_{NN}-V_{nNNN}$=1.5 eV were somewhat exagerated, these interactions certainly play an important role.\\
(iv) the magnetic coupling and hopping integral between adjacent atoms depend in a stereospecific manner
on the existence and position of hole(s) in the immediate vicinity. A simple correlation with the electrostatic
energies of the intermediate ligand to metal charge transfer states has been proposed resulting in simple
formulas, which should be used in a realistic model Hamiltonian. From the exploratory calculations appearing
in Table 5, and for sake of simplicity, it seems sufficient to consider the first neighbors of the bond
in the calculation of the electrostatic energy changes $\delta E_{CT1}+\delta E_{CT2}$ and 
$\delta E_{CT}-\delta E_0$ appearing in formulas (27) and (28), respectively.

One may therefore propose the following extended t-J model Hamiltonian:
\begin{eqnarray}
H&=&\sum_{<pq>}^{NN} t(\delta E)\cdot (a_p^{\dag} a_q + a_q^{\dag}a_p + s. c.) \delta (n_p + n_q, 1) - J(\delta E) \cdot (S_p \cdot S_q)+\nonumber\\
&+&\sum_{<pr>}^{NNN} t_{NNN} \cdot (a_p^{\dag} a_r + a_r^{\dag}a_p + s. c.) \delta (n_p + n_r, 1) -J_{NNN} \cdot (S_p\cdot S_r)+\nonumber  \\   
&+&\sum_{<pqrs>}^{plaquette} K \cdot [ (S_p\cdot S_q)(S_r\cdot S_s)+(S_p\cdot S_s)
             (S_q\cdot S_r)-(S_p\cdot S_r)(S_q\cdot S_s)]+\nonumber \\ 
&+&\sum_{<pqr>}^{connected} h_{SD}^{pqr} \cdot (a_{p}^{\dag}a_{\bar{q}}^{\dag}a_{q}a_{\bar{r}}+
a_{r}^{\dag}a_{\bar{q}}^{\dag}a_q a_{\bar{p}} - a_{p}^{\dag}a_{\bar{q}}^{\dag}a_{r}a_{\bar{q}}
-a_{r}^{\dag}a_{\bar{q}}^{\dag}a_p a_{\bar{q}}+ s.c.) \delta(n_p+n_q+n_r,2)+\nonumber \\
&+&\sum_{pq}^{\le nNNN} V_{pq} \cdot \delta(n_p,0) \cdot \delta(n_q,0) \nonumber 
\end{eqnarray}
In this Hamiltonian, $t(\delta E)$ and $J(\delta E)$ reflect the dependence of the first-neighbor interactions
on the number and position of the adjacent holes, which is controlled by the energy changes in the
charge transfer intermediates. The second-neighbor interactions are not influenced by the presence of 
adjacent holes and the displacement of a singlet takes placed 
toward an adjacent hole, so the positions occupied by the singlet and the hole have to be connected.

The here-proposed modifications with respect to the usual t-J or t-J-V Hamiltonians are important. We 
would like to point out that some local physical effects evidenced in the present work may have an
impact on the spatial ordering of charges an spins and on their dynamics. In a crude static look at 
this problem one may notice that stripping separation of charges and spins 
(i) favors the mobility of holes in the charged column, since $t$ and $J$ values become
especially larger, (ii) leads to a stronger magnetic coupling in the bonds perpendicular to the
stripes ( with a trend to form singlets on these bonds). The two last effects would result from
the three body corrections on $t$ and $J$ created by the adjacent holes. Finally, (iii) the mobility
of the stripes can be enhanced by the second-neighbor hopping integral and by the singlet displacement
operator. 

We think that our calculations are reliable enough and that their qualitative conclusions and
quantitative estimates  deserve
to be considered in simulations of
the collective properties of the lattice, whatever the method used for such a study.

\section*{Acknowledgements}
The authors wish to thank Daniel Maynau for computational help and development.
The ROHF calculations were done using the MOLCAS package\cite{MOLCAS}, the DDCI calculations
were done using the CASDI package \cite{CASDI}. 
The authors are indebted to the European Commission for the TMR network contract
 ERBFMRX-CT96-0079, Quantum Chemistry of Excited States.  C.J.C. acknowledges
 the financial support through the TMR activity "Marie Curie research training 
grants" Grant No. HPMF-CT-1999-00285 established by the European Community.

\bibliographystyle{aip}

\newpage

\section{Figure captions}
Figure 1. Summary of the strategy used to extract effective interactions in La$_{2-x}$Sr$_x$CuO$_4$ systems.
\\
Figure 2. Representation of the clusters employed in the ab initio calculations. (a) the plaquette Cu$_4$O$_{12}$,
(b) the linear cluster Cu$_3$O$_{10}$. The first-neighbors of the cluster atoms have been also included.
They have been modeled by means of the combination of a pseudopotential and a point charge, to mimic the
both the exclusion and the Coulombic effects.
\\
Figure 3. Localized active orbitals for the plaquette: (a) for undoped systems, centered
in 3$d_{x^2-y^2}$, with important tails on the four in-plane neighbor oxygen atoms. (b) for hole-doped systems a
strong $3d-2p$ rehybridization takes places, the $2p$ character increases sustantially with respect to the undoped situation.
\\
Figure 4. Linear correlation between the calculated first-neighbor interactions in binuclear clusters and the change of the electrostatic energy in the intermediate charge transfer states (in eV): (a) $J_{NN}$ versus 
$\delta E_{CT1}+ \delta E_{CT2}$; (b) $t_{NN}$ versus  
$\delta E_{CT}- \delta E_0$.
\\

\newpage
\begin{figure}[htb]
\vbox to 20.0cm{
\includegraphics{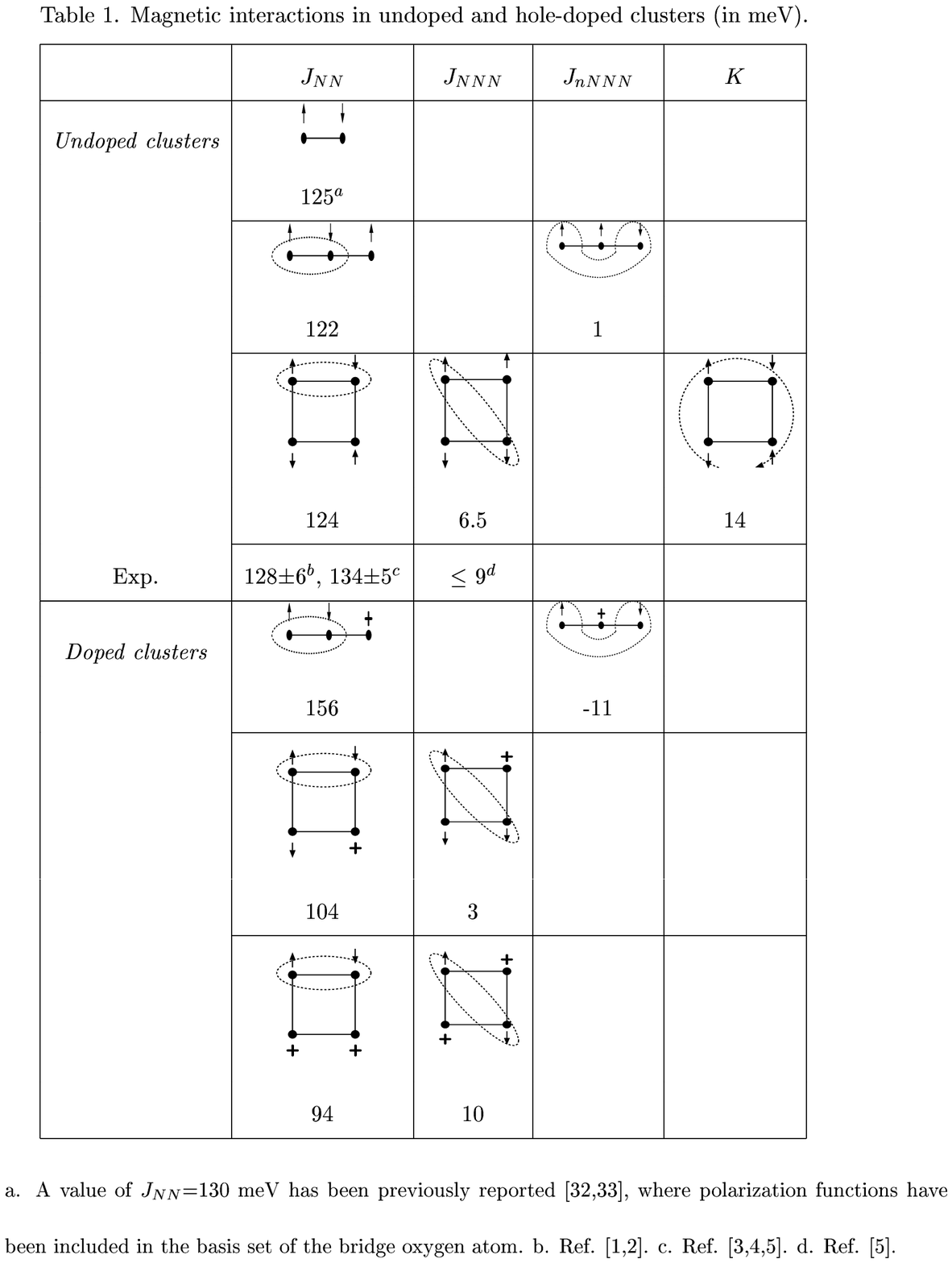}
\vfill}
\end{figure}  
\newpage

\begin{figure}[htb]
\vbox to 20.0cm{
\includegraphics{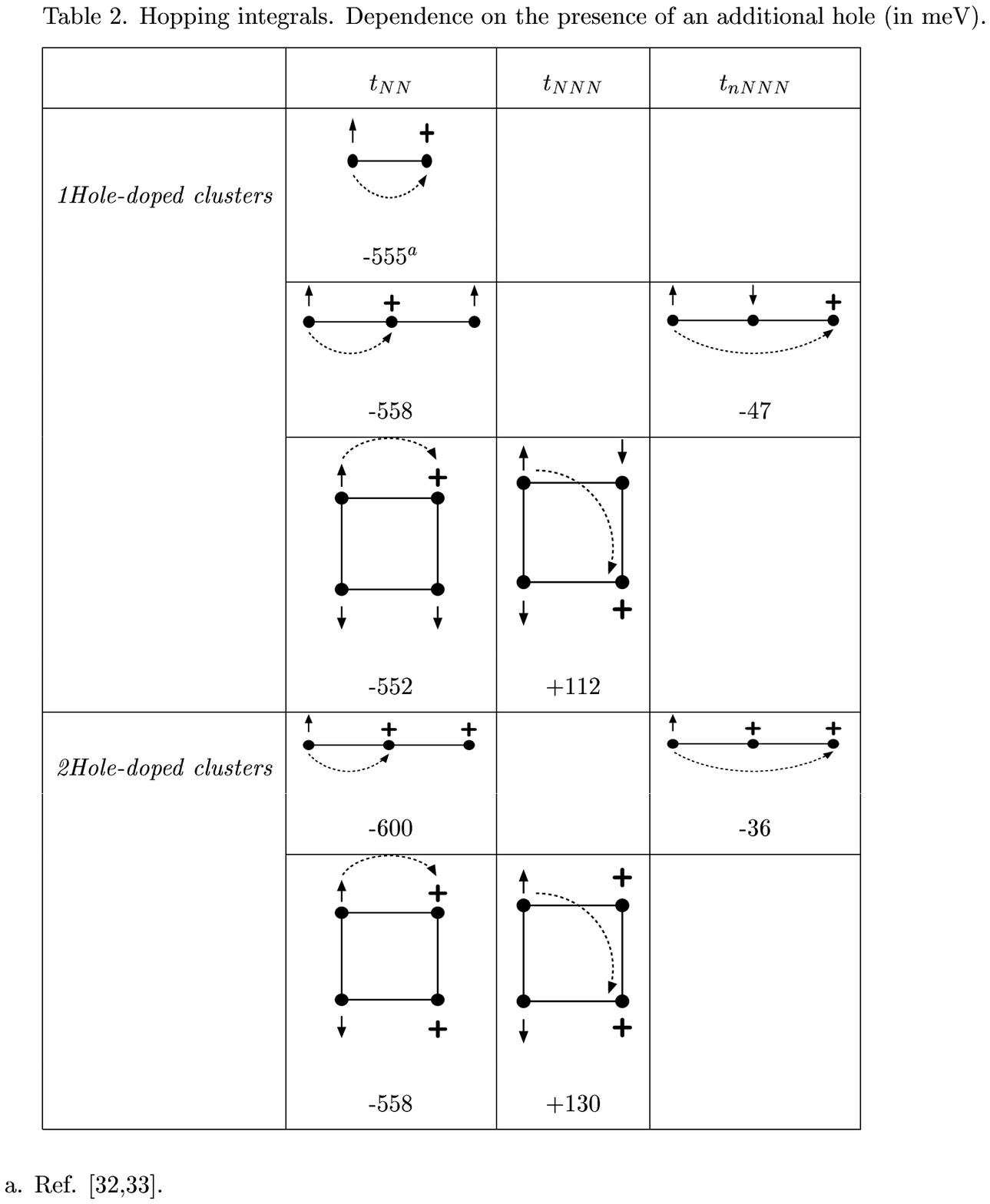}
\vfill}
\end{figure}  
\newpage

\begin{figure}[htb]
\vbox to 20.0cm{
\includegraphics{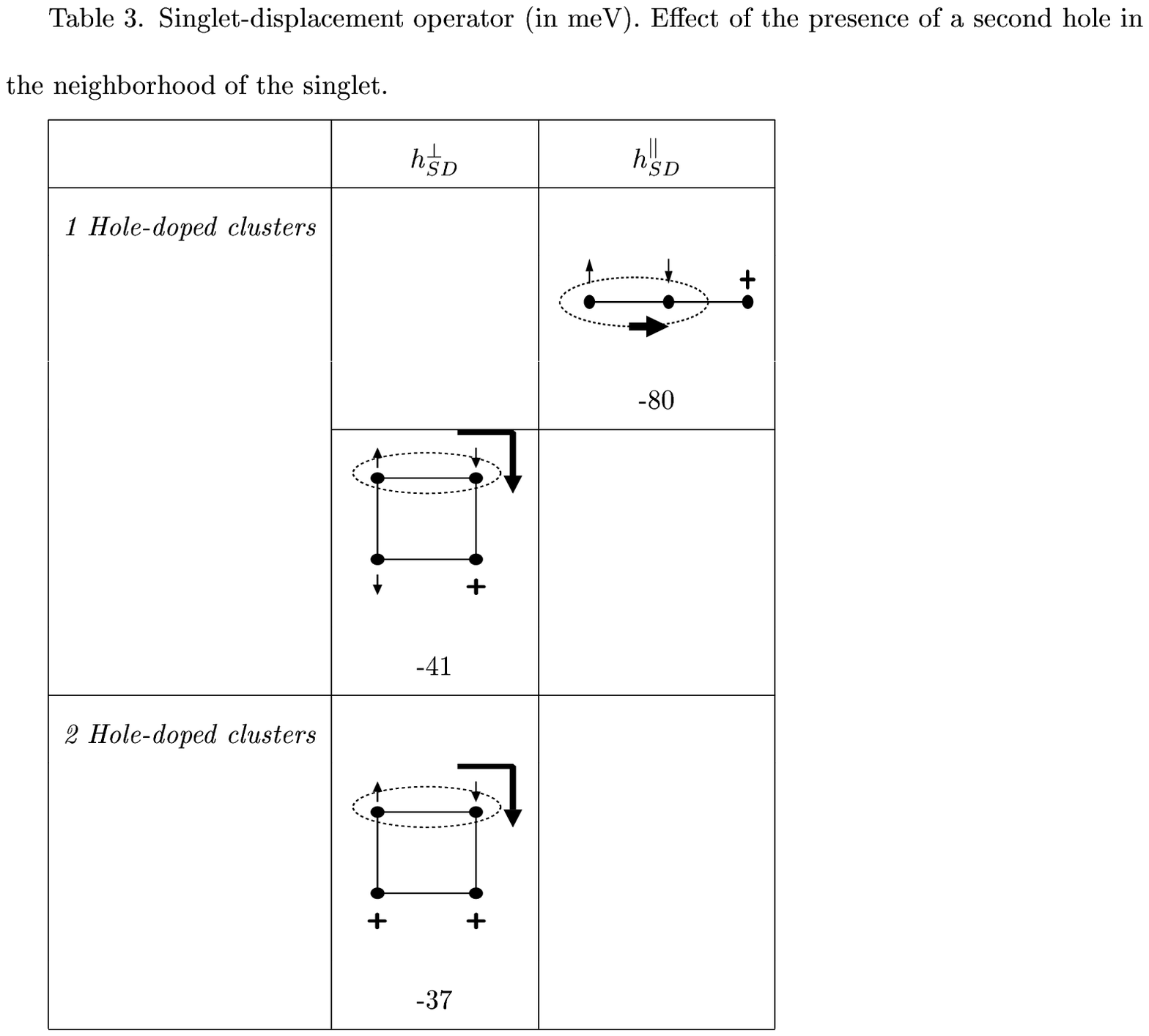}
\vfill}
\end{figure}  
\newpage

\begin{figure}[htb]
\vbox to 20.0cm{
\includegraphics{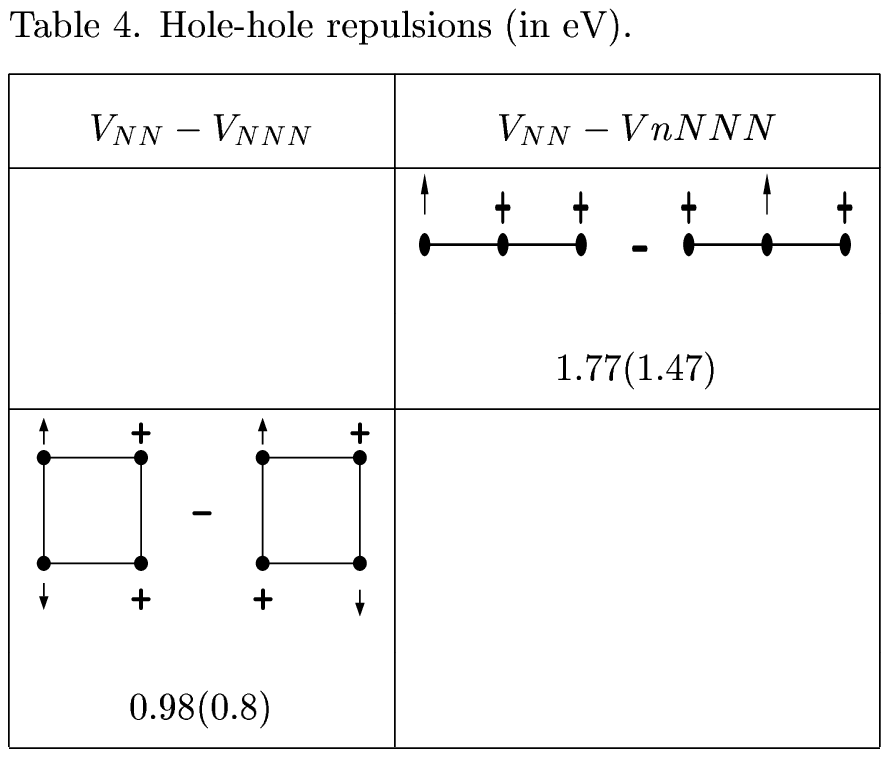}
\vfill}
\end{figure}  
\newpage

\begin{figure}[htb]
\vbox to 20.0cm{
\includegraphics{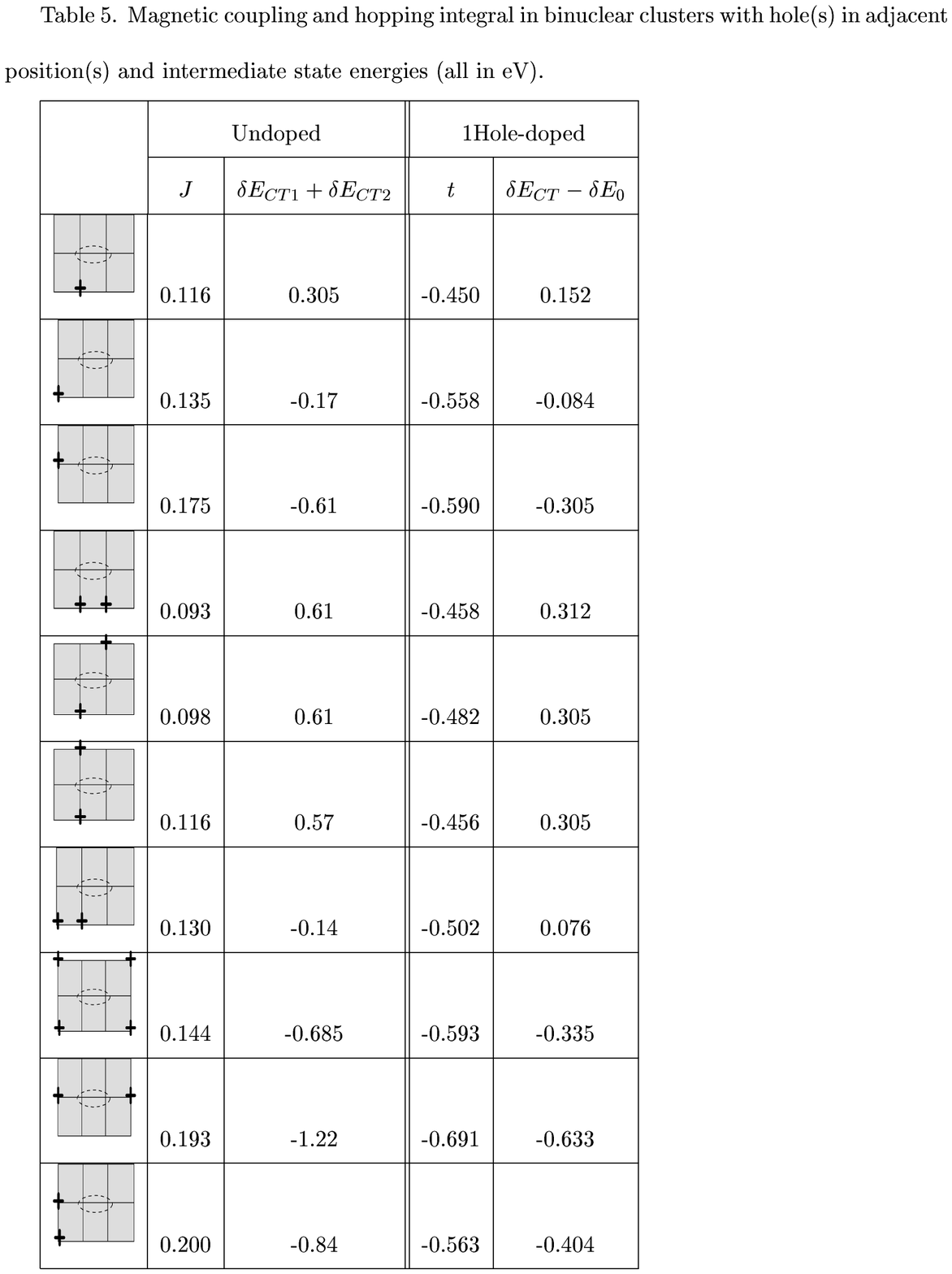}
\vfill}
\end{figure}  
\newpage

\begin{figure}[htb]
\vbox to 14.0cm{
\includegraphics{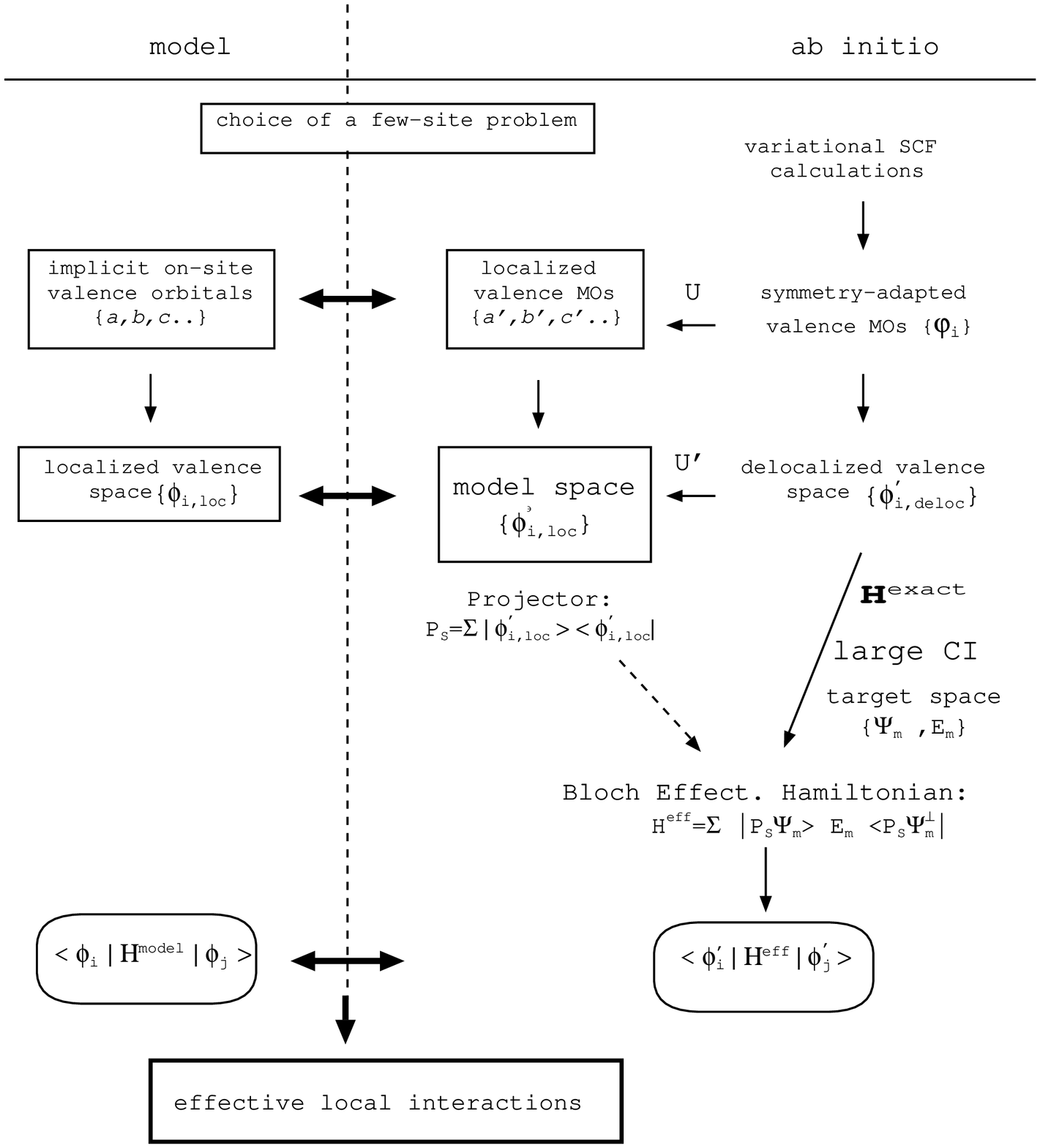}
\vfill}
Figure 1. Calzado and Malrieu
\end{figure}  

\newpage
\begin{figure}[htb]
\vbox to 14.0cm{
\includegraphics{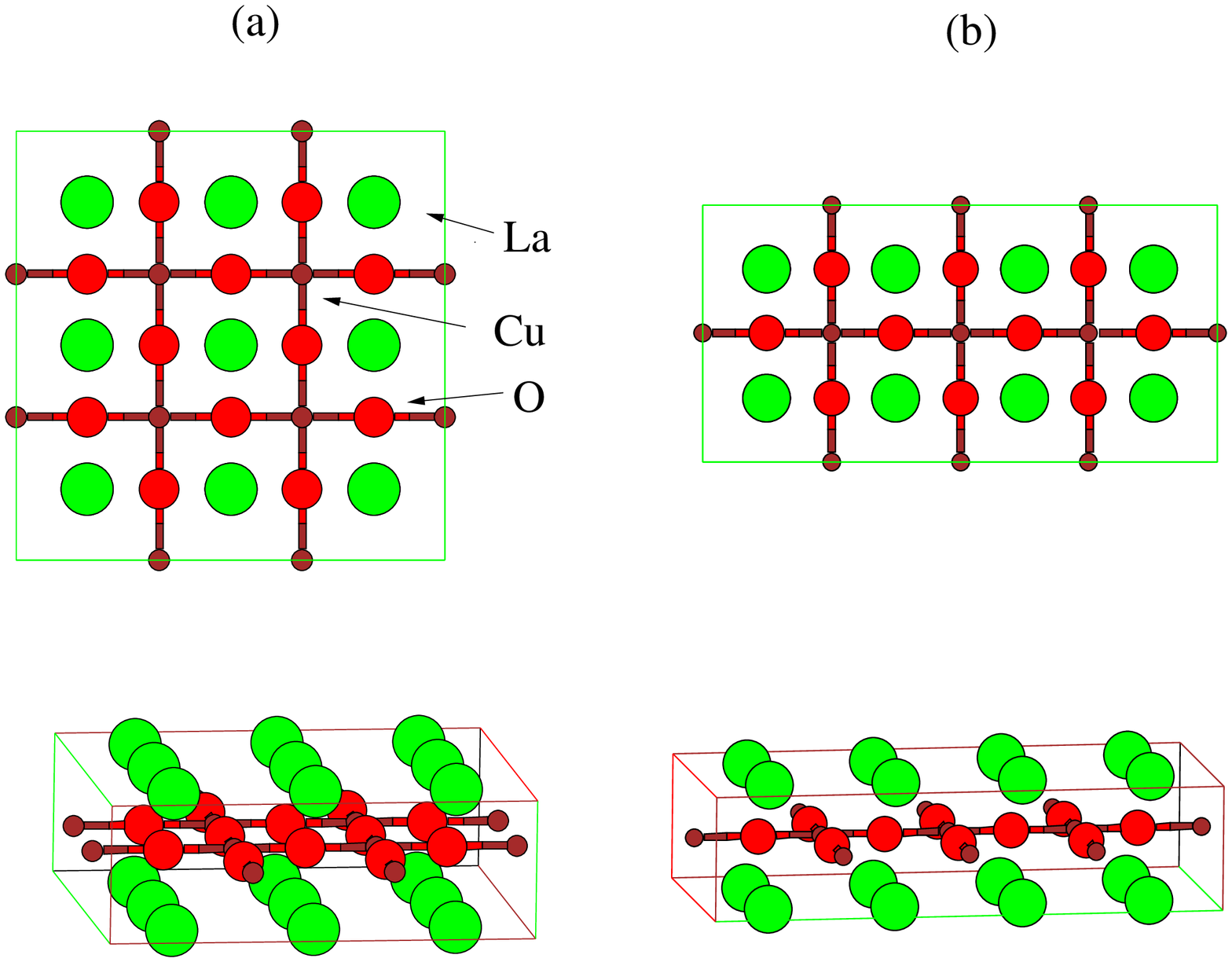}
\vfill}
Figure 2. Calzado and Malrieu
\end{figure}  

\newpage
\begin{figure}[htb]
\vbox to 14.0cm{
\includegraphics{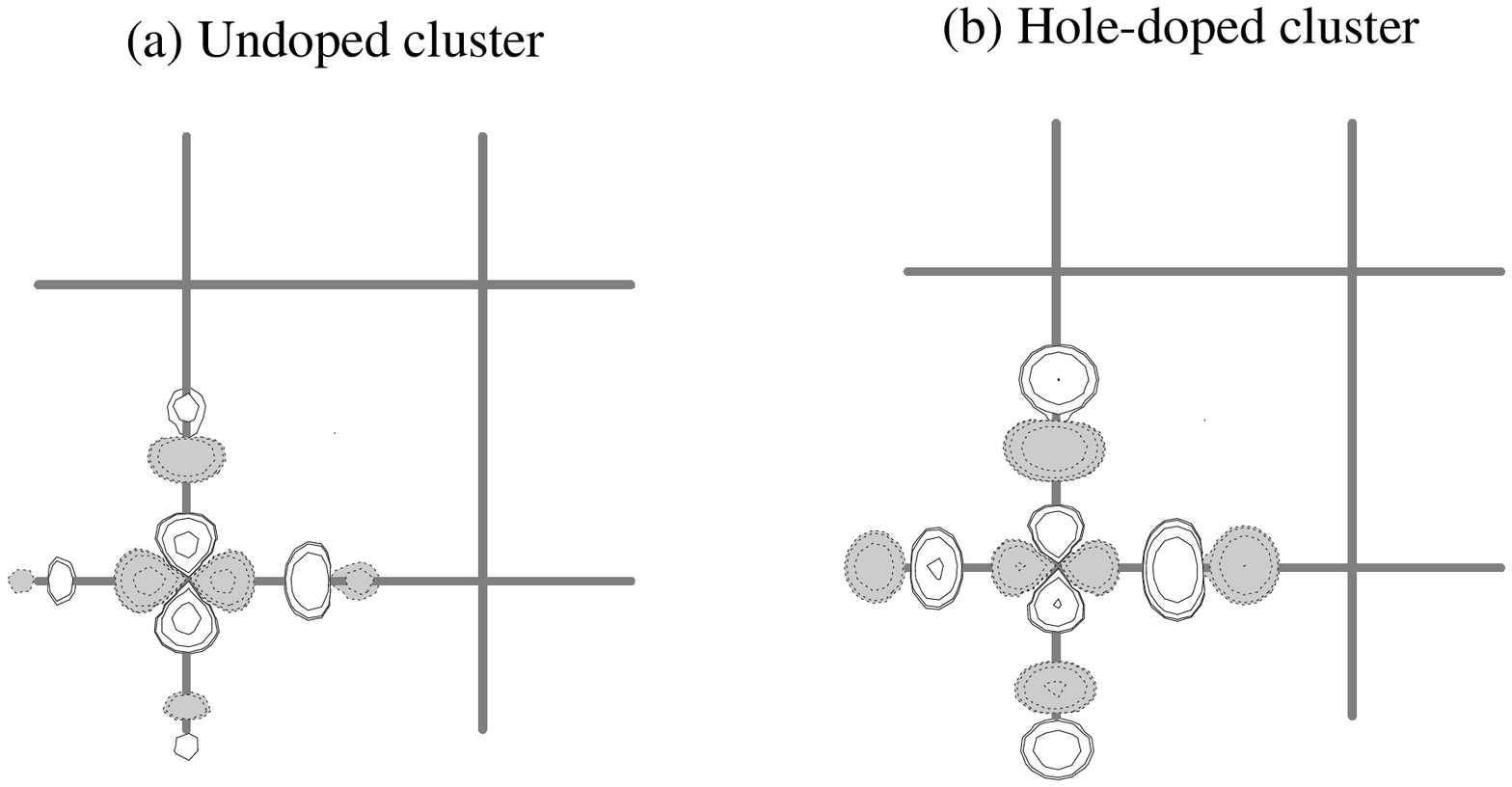}
\vfill}
Figure 3. Calzado and Malrieu
\end{figure}  

\newpage
\begin{figure}[htb]
\vbox to 14.0cm{
\includegraphics{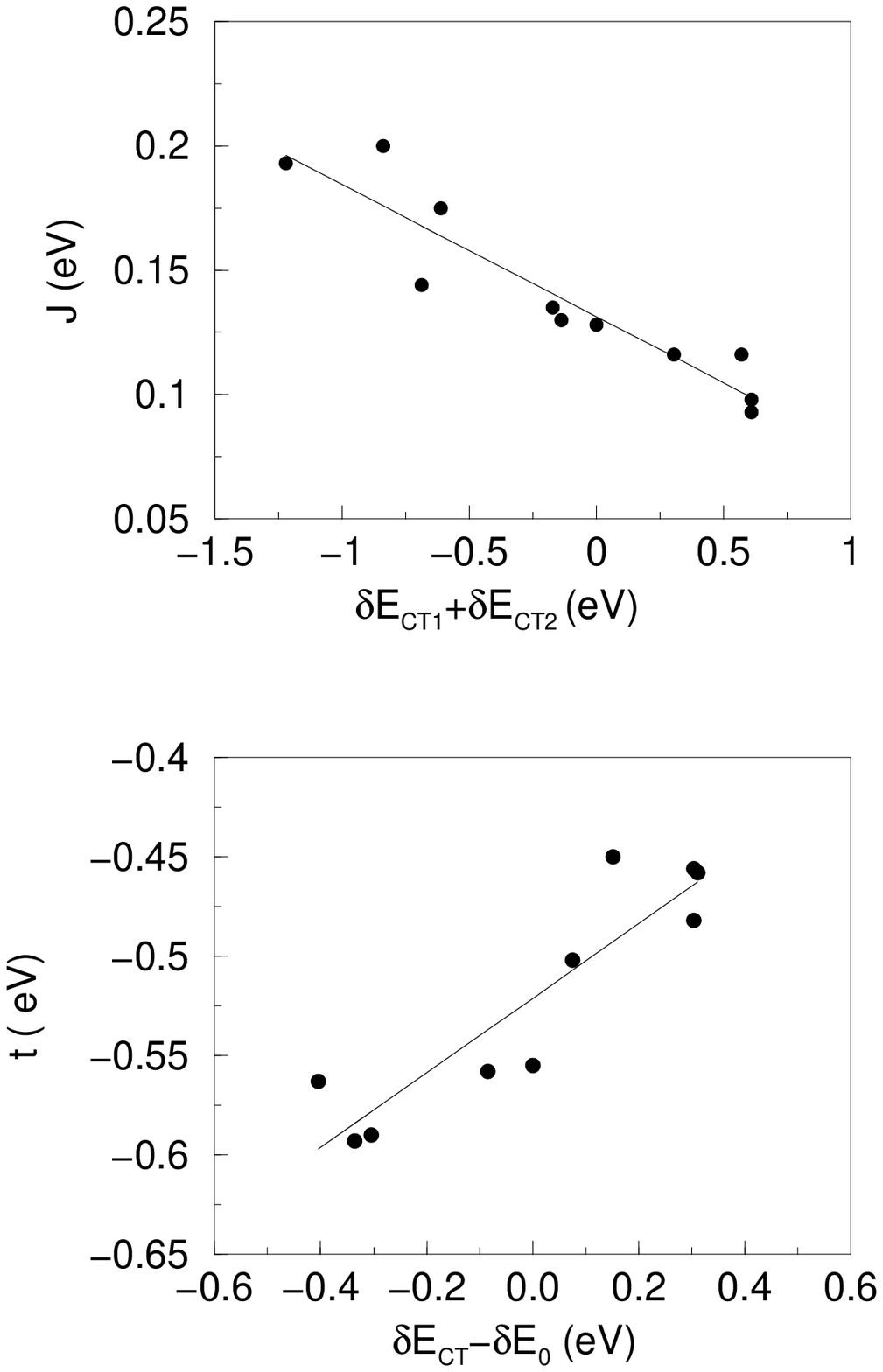}
\vfill}
Figure 4. Calzado and Malrieu
\end{figure}  
\end{document}